\documentclass[12pt]{iopart}

\expandafter\let\csname equation*\endcsname\relax
\expandafter\let\csname endequation*\endcsname\relax
\usepackage{amsmath} 
\usepackage{graphicx}
\usepackage{subfig}
\usepackage{float}
\usepackage{CJKutf8}

\usepackage[square, comma, sort&compress, numbers]{natbib}
\begin{document}
\begin{CJK*}{UTF8}{gbsn}

\title{Quasi-single-stage optimization for permanent magnet stellarators}

\author{Guodong Yu$^1$, Ke Liu$^1$, Tianyi Qian$^2$, Yidong Xie$^1$, Xianyi Nie$^1$, Caoxiang Zhu$^{1,*}$}
\address{$^1$ School of Nuclear Science and Technology, University of Science and Technology of China, Hefei 230027, China}
\address{$^2$ Princeton University, Princeton, New Jersey 08544, USA}

\ead{caoxiangzhu@ustc.edu.cn}
\vspace{10pt}

\begin{indented}
\item[]January 2024
\end{indented}
\end{CJK*}

\begin{abstract}
Advanced stellarators are typically optimized in two stages.
The plasma equilibrium is optimized first, followed by the design of coils/permanent magnets.
However, the coils/permanent magnets in the second stage may become too complex to achieve
the desired equilibrium.
To address this problem, a quasi-single-stage optimization method has been proposed. 
In this paper, we introduce this method for designing permanent magnet (PM) stellarators.
The new approach combines straightforward PM metrics to penalize the maximum required PM thickness and the 
mismatch between the fixed-boundary equilibrium and the free-boundary one, along with typical physical 
targets. Since the degrees of freedom of the PMs are not included and directly used to minimize the objective 
function in this method, so we call it ``quasi-single-stage" optimization. 
We apply this quasi-single-stage optimization method to find a new quasi-axisymmetric PM design.
The new design starts from MUSE, which was initially designed using a two-stage optimization approach.
The resulting design, \mbox{MUSE++}, exhibits an order of magnitude lower quasi-symmetric error and a 
one-order reduction in normal field error. 
 We show that \mbox{MUSE++} has approximately 30\% fewer magnets compared to a proxy model ``\mbox{MUSE-0}"
 that uses the same FAMUS optimization without the benefit of a single-stage equilibrium optimization.
These results demonstrate that the new single-stage optimization method can concurrently
improve plasma properties and simplify permanent magnet complexity.

\end{abstract}


\section{Introduction} 
\label{section:Introduction}
Stellarators are three-dimensional (3D) magnetic confinement fusion devices to accommodate
high-temperature plasmas without driving plasma currents \cite{Spitzer1958The}. 
Stellarators were projected to have worse confinement than tokamaks and thus 
encountered a long period of overlook. The concept of advanced stellarators 
opens up a new path for the development of stellarators \cite{Horst1999Theory}. 
The remarkable experimental records of W7-X prove that meticulously optimized 
stellarators possess as good confinement performance as tokamaks at the same 
experimental scale \cite{Stroth2021Stellarator-tokamak}. Stellarators have 
become an attractive path to fusion energy due to the advantages like being 
steady-state, disruption-free, and high-density operation, etc.

To design advanced stellarators, the so-called ``two-stage" optimization approach \cite{NELSON2003Design} 
is widely used. In the first stage, the plasma equilibrium is optimized until the physical properties, 
such as good flux surfaces, magnetohydrodynamic (MHD) stabilities, neoclassical transport, etc.,
 meet the goals. In the second stage, 3D-shaped coils are designed such that the desired equilibrium 
 can be produced. There are different coil designing codes, e.g. the surface current method code 
 NESCOIL \cite{Merkel1987Solution} and the direct nonlinear optimization code FOCUS \cite{Zhu2017New}. 
 However, the coils designed in the second stage might be too complicated to be easily fabricated. 
 In that case, one would have to return to the first stage and adjust the target plasma equilibrium. 
 This loop might need to be done several times until a satisfying pair of plasma and coils is found.  

To improve efficiency and robustness, a single-stage optimization approach that can optimize plasma and coils simultaneously has been explored. In general, single-stage optimization approaches can be divided into two types. The first one starts from the plasma equilibrium and the parameters to describe 
fixed-boundary equilibria are varied. During the optimization, figures of merit to assess coil complexities 
are added \cite{Henneberg2021Combined, Wechsung2022Single, Jorge2023Single}. The other type starts from coils. 
The coil shapes are varied and the physical properties are evaluated from free-boundary equilibria or vacuum 
fields \cite{Yu2021A, Yu2022Existence, Giuliani2022Single, Giuliani2022Direct}. 

Permanent magnet (PM) is an attainable alternative to electromagnetic coils. A closed toroidal flux surface can be created by a suitable layout of PMs together with toroidal field (TF)
 coils \cite{Helander2020Stellarators}. Recently, several methods have been developed to design PMs. 
 Initial explorations employ linear methods. The distribution of magnetic dipoles can be derived from 
 the surface current potential \cite{Zhu2020Designing}, which we shall call it the ``surface current method'' 
 hereafter. The thickness and polarization of permanent magnets can also be optimized by a least-squares 
 minimization method \cite{Landreman2021Calculation}. Later methods optimize PMs 
 individually \cite{Zhu2020Topology, Xu2021Design, Lu2021Design, Kaptanoglu2023Greedy} to include more 
 realistic constraints, like in uniform shapes/strength and having limited orientations. Discrete small 
 cubic PMs are arranged in advance at specific locations. Magnetization orientations are optimized globally using a topology optimization method \cite{Zhu2020Topology} or locally chosen from a set of available polarization using greedy algorithms \cite{Lu2022Development, Kaptanoglu2022Permanent, Kaptanoglu2023Greedy}.
  All the methods assume a target plasma equilibrium which is fixed during the optimization of permanent magnets. 
  In other words, PM stellarators, like PM4Stell \cite{Zhu2022PM4Stell, Hammond2022Design} and MUSE \cite{Qian2022Simpler, Qian2023Design}, are based on the two-stage optimization approach and therefore face similar difficulties. Some equilibria might require numerous PMs or cannot be realized by available PMs.

In this work, we develop a new kind of optimization approach for PM stellarators, 
in which the target plasma equilibrium and PMs are optimized simultaneously. 
We follow the typical fixed-boundary optimization for plasma equilibria. 
In addition to common objective functions for the equilibrium, 
we add a simple metric to penalize the maximum required PM thickness and the mismatch between the fixed-boundary equilibrium and the free-boundary one. This method belongs to the first type of single-stage optimization approach, where the degrees of freedom are the fixed-boundary equilibrium coefficients. However, our method is not a complete ``single-stage''  optimization, as we are using a proxy function during the optimization of the plasma boundary and a follow-up process is still required to design the actual magnets, just like the second stage in the ``two-stage'' approach.
For this reason, we call our new method the ``quasi-single-stage optimization ''.
The proxy function comes from the surface current potential and is similar to the coil complexity metrics employed
in previous studies \cite{Drevlak2019Optimisation, Carlton-Jones2021Computing}. 
We apply our quasi-single-stage optimization approach to obtain a new quasi-axisymmetric PM stellarator. 
It starts from MUSE \cite{Qian2022Simpler, Qian2023Design}. MUSE was originally designed with the two-stage 
optimization approach \cite{Qian2022Simpler} and is the first completely constructed PM stellarator 
in the world \cite{Qian2023Design}. 

This paper is organized as follows. In section \ref{section:PM}, we give a brief introduction to the PMs' 
design methods related to this work. In section \ref{section:Single-stage}, details of the new single-stage 
optimization approach are introduced. Numerical results for the improved MUSE case are described in 
section \ref{section:Numerical}. Realistic PM designs are presented in section \ref{section:Improved}. 
Finally, section \ref{section:Discussion} gives a summary. 

\section{PM design methods}
\label{section:PM}
There are different design methods for PMs. The linear methods are fast but have few controls in the locations and/or 
orientations of PMs. Nonlinear methods are more sophisticated and can obtain more practical solutions. 
They are also much slower and would be computationally expensive if included in single-stage optimizations. 
Therefore, we use a metric derived from the surface current method to quickly approximate the total number of 
permanent magnets required. With this metric, quasi-single-stage optimization is performed. Afterward, we use the PM 
optimization code, FAMUS \cite{Zhu2020Topology}, to obtain a practical design for PMs and compare it with the 
reference designs. In the following, we briefly introduce the surface current method and FAMUS.

\subsection{The surface current method}
The surface current method comes from the original modular coil design method.
First implemented in the NESCOIL code \cite{Merkel1987Solution}, the method assumes that there is a closed toroidal 
winding surface wrapping a target plasma and calculates the surface current on the winding surface to produce the 
desired magnetic field. The surface current density $\mathbf{K}$ can be represented as
\begin{equation}
\mathbf{K}=\mathbf{n}' \times \nabla{\Phi} \ ,
\end{equation}
where $\mathbf{n}'$ is the normal vector of the winding surface and $\Phi$ is the surface current potential. Hereafter, 
the quantities with primes are associated with the winding surface with unprimed quantities indicating the plasma surface.  
The current potential $\Phi$ is defined 
as $\Phi(\theta^\prime,\zeta^\prime)=\Phi_{sv}(\theta^\prime,\zeta^\prime)+\frac{G\zeta^\prime}{2\pi}+\frac{I\theta^\prime}{2\pi}$, 
where $\Phi_{sv}$ is single-valued, $G$ and $I$ are the currents that link plasma poloidally and toroidally.

To calculate the surface current, the normal component of the magnetic field $\mathbf{B}$ on the last closed flux 
surface (LCFS) is minimized,
\begin{equation}
\chi^2_B=\iint_S(\mathbf{B}\cdot \mathbf{n})^2 \mathrm{d}S \ ,
\end{equation}
where $S$ represents the plasma surface and $\mathbf{n}$ the unit surface normal vector.
The magnetic field $\mathbf{B}$ may have various contributions, $\mathbf{B}=\mathbf{B}_{K} + \mathbf{B}_{external} +\mathbf{B}_{plasma}$, where $\mathbf{B}_{K}$ is generated by the surface current on the winding surface, $\mathbf{B}_{external}$ from the fixed external coils (such as TF coils), and $\mathbf{B}_{plasma}$ from the plasma currents. Giving the plasma surface, plasma currents, external field, G, I, and the winding surface, the problem becomes a least-squares minimization and can be linearly solved. The REGCOIL code \cite{Landreman2017An} improves NESCOIL by adding a Tikhonov regularization and the objective function becomes
\begin{equation}
\chi^2=\chi^2_B + \lambda \chi^2_K \ ,
\end{equation}
where $\chi^2_K =\int |\mathbf{K}(\theta^\prime,\zeta^\prime)|^2 dS^\prime$ is the surface-average-squared 
current density and $\lambda$ the regularization parameter. In this paper, we use REGCOIL to calculate the 
surface current.

The surface current potential can be discretized into localized magnetic dipoles, whose magnetization orientations 
are along the normal directions of the surface, and the magnetization $\mathbf{M}$ can be expressed with a 
Dirac delta function, $\mathbf{M}=\Phi\mathbf{n}' \delta(s-s_0)$, where $s_0$ labels the winding surface. 
Then the magnetic moment of each dipole is computed as,
\begin{equation} \label{eq:dipole}
\mathbf{m}=\Phi \mathbf{n}' \Delta S' \ ,
\end{equation}
where $\Delta S'$ is the area of winding surface element. This method can be further extended to design PM with multiple layers. Details can be found in Ref. \cite{Zhu2020Designing}.
Following equation \eqref{eq:dipole}, we can use the current potential $\Phi$ to approximate the characteristics of PM. For example, the thickness of the required PM is correlated with $\Phi$.

\subsection{PM optimization using FAMUS} 
To initialize a FAMUS run, one must provide geometry information for all magnets.
This could be done by using the MAGPIE code \cite{Hammond2020Geometric} or simply using structured grids.
Since the magnet size is generally small (large ones can also be decomposed into small ones), each magnet 
is represented by a magnetic dipole and a local spherical coordinate is used to indicate the polarization,
\begin{equation}
    \mathbf{m}(p, \theta, \phi) = p^q m_0 \begin{pmatrix}
    \sin{\theta} \cos{\phi} \\
    \sin{\theta} \sin{\phi} \\
    \cos{\theta}  \end{pmatrix} \ ,
\end{equation}
where $p\in[0,1]$ is the free parameter controlling the magnitude, $q$ the user-defined penalization 
coefficient, $m_0$ the maximum allowable magnetic moment, $\theta \in [0, \pi]$ the polar angle, 
and $\phi \in [-\pi, \pi)$ the azimuth angle. The maximum allowable magnetic moment $m_0 = \iiint M \mathrm{d}V$ ($V$ 
the volume of the magnet) is determined by the material magnetization $M$. In this paper, 
we use $M=1.165\times10^6 \mathrm{A/m}$.

FAMUS has several objective functions. The primary one is the normal field error, which is similar to $\chi^2_B$ in 
the surface current method,
\begin{equation}
f_B=\frac{1}{2}\iint_S[\mathbf{B}_{PM} \cdot \mathbf{n} - B_n^{target}]^2dS \ ,
\end{equation}
where $\mathbf{B}_{PM}$ is the total magnetic field from PMs and $B_n^{target}$ is the target normal field to be canceled.
If the plasma boundary is one of the flux surfaces, we have $B_n^{target} =-(\mathbf{B}_{coils}+\mathbf{B}_{plasma})\cdot\mathbf{n}$, where $\mathbf{B}_{coils}$ comes from external coils and $\mathbf{B}_{plasma}$ from plasma currents. 
In addition to $f_B$, there is also another objective function that is often used,
\begin{equation}
f_D=\sum|p^q|(1-|p^q|) \ .
\label{eq:f_D}
\end{equation}
$f_D$ will penalize intermediate values for the normalized magnitude $p^q$ to achieve binary distributions 
(either $0$ or $\pm 1$).
The final cost function in FAMUS is the weighted summation of each objective function. FAMUS then employs 
a quasi-Newton method to find the minimum of the final cost function with $\{p, \theta, \zeta\}$ of each 
dipole being the degrees of freedom.
After optimization, magnets with almost zero magnitudes will be removed.
FAMUS solutions can be further improved to discrete orientations \cite{Hammond2022Design}.

\section{Quasi-single-stage optimization for PM}
\label{section:Single-stage}
The quasi-single-stage optimization method for PM proposed in this paper is to add PM metrics to the conventional 
fixed-boundary optimization loop.
The fixed-boundary optimization tool used in this paper is SIMSOPT \cite{Landreman2021simsopt} with 
VMEC \cite{Hirshman1983steepest} being the equilibrium code.
The independent variables for optimization are the Fourier amplitudes of boundary flux 
surface ${R_{mnc},Z_{mns}}$, and in the cylindrical coordinates, boundary flux surface with stellarator 
symmetry is defined as,
\begin{align}
R(\theta, \phi)=\sum_{m,n}R_{mnc} \cos(m\theta-n_{fp}n\phi) \ ,\\
Z(\theta, \phi)=\sum_{m,n}Z_{mns} \sin(m\theta-n_{fp}n\phi) \ ,
\end{align}
where $\phi$ is the standard cylindrical angle, $\theta$ is the poloidal angle, and $n_{fp}$ is the number of toroidal field periods.
The boundary shape is varied to minimize a total objective function that consists of plasma metrics and PM metrics,
\begin{equation}
f_{total}=f_{plasma} + f_{PM} \ .
\end{equation}

\subsection{Plasma metrics}
The optimization targets for the plasma include quasi-axisymmetry, rotational transform, 
and volume inside the boundary surface. Our main goal is to further improve quasi-axisymmetry 
while constraining the rotational transform and the plasma volume. 
So, the objective function for plasma equilibrium is  
\begin{equation}
f_{plasma}=\omega_{QA} f_{QA} + \omega_{\iota} f_{\iota} + \omega_{V} f_{V} \ ,
\end{equation}
where $\omega_i$ is the user-provided weight associated with the objective term $f_i$. $f_{QA}$ is the normalized symmetry-breaking modes of magnetic field $B$, given by, 
\begin{equation}
f_{QA}=\sum_{s_j}\left(\sum_{m,n\neq 0} \left( \frac{B_{m,n}}{B_{0,0}}\right)^2\right) \ ,
\end{equation}
where $B(s,\theta,\zeta)=\sum_{m,n} B_{m,n} \cos(m\theta_B - n_{fp} n\zeta_B)$, $s$ labels the normalized  toroidal flux, $j$ the flux number, $\theta_B$ and $\zeta_B$ the poloidal and toroidal angle in Boozer coordinates \cite{Boozer1981Plasma}. The Boozer coordinate transformation and the Fourier decomposition of $B$ are done by the BOOZ\_XFORM code \cite{Boozer1981Plasma}.
The rotational transform objective function,
\begin{equation}
f_\iota = (\bar{\iota}-\bar{\iota^*})^2 \ ,
\end{equation}
where $\bar{\iota} = \int_0^1\iota ds$ is the average rotational transform.
Here, we use $\bar{\iota^*}=0.195$ and it is slightly smaller than $0.2$ to avoid low-order rational surfaces. 
The plasma volume objective function is
\begin{equation}
f_V = (V-V^*)^2 \ ,
\end{equation}
where $V$ is the plasma volume and $V^*$ is the target value.
In this paper, we use $V^*=0.012 \ \mathrm{m^3}$.
The major radius and the number of toroidal field periodicities are fixed during the optimization.

\subsection{PM metrics}
Inside each iteration, REGCOIL is used to solve the surface current potential.
In this paper, we fix the winding surface to be a circular torus with a major radius of 30 cm and a minor 
radius of 10 cm. The toroidal field is produced by an infinite long wire with a specific current $I_{TF}$ 
placed in the center of the torus.
Plasma currents are zero since we set the plasma pressure to zero.
The toroidally and poloidally linked currents are also zero $G=I=0$.
We ran a scan for the regularization parameter $\lambda$ with 30 values in the range $10^{-24} \sim 10^{-13}$.
The optimal value $\lambda_0$ is chosen to have the lowest value for $\chi^2_B\cdot \chi^2_K$.

With the current potential $\Phi$ at $\lambda_0$, we obtain two metrics for PM,
\begin{equation}
f_{PM}=\omega_{\Phi} f_{\Phi} + \omega_{\chi^2_B} f_{\chi^2_B} \ . 
\end{equation}
The first term,
\begin{equation}
f_{\Phi}=(\Phi_{max}(\lambda_{0}))^2 \ ,
\end{equation}
penalizes the maximum value of the current potential, which is a proxy for the largest required thickness for PMs.
The second term,
\begin{equation}
f_{\chi^2_B}=(\chi^2_B(\lambda_{0}))^2 \ ,
\end{equation}
penalizes the residual normal field error from surface currents.
Reducing $\chi^2_B$ will avoid possible mismatches between the fixed-boundary target and the free-boundary 
equilibria generated by continuous surface currents. 

\subsection{Optimization process}
\begin{figure}[htbp]
    \centering
    \includegraphics[scale=0.18]{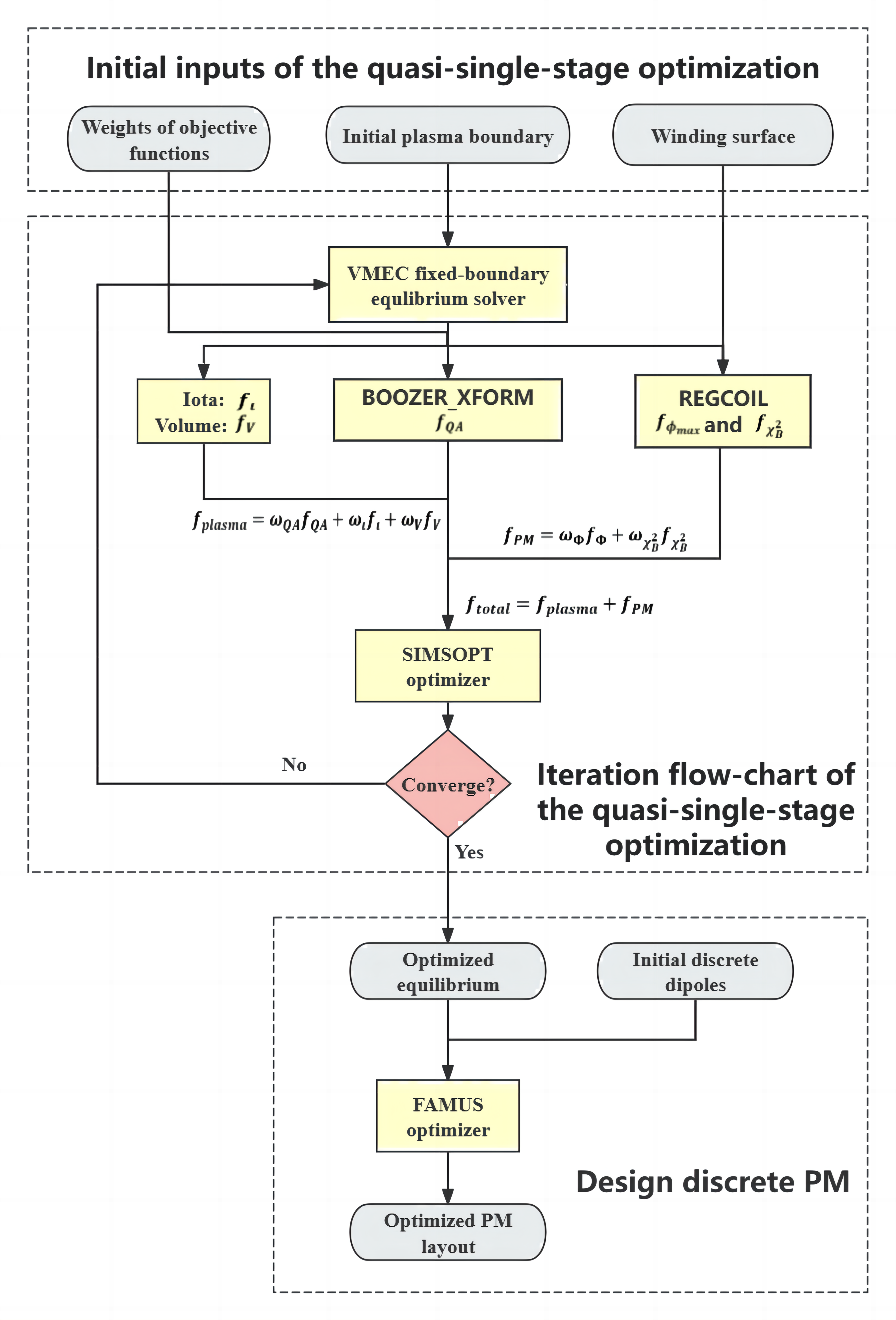}
    \caption{Iteration flow-chart of the quasi-single-stage optimization algorithms.}
    \label{fig:flow_chart}
\end{figure}

The flowchart of the complete optimization process is shown in figure \ref{fig:flow_chart}.
The input consists of the initial plasma boundary, a pre-described winding surface used for REGCOIL, and weights $\omega_i$.
The VMEC code will calculate the fixed-boundary equilibrium and then the quantities, iota and plasma volume, are then obtained.
The BOOZER\_XFORM code is called to compute the Fourier modes of $|\mathbf{B}|$ in the Boozer coordinates to calculate the QA quality.
REGCOIL is also called to compute the surface current potential on the given winding surface, then $f_{\Phi}$ and $f_{\chi^2_B}$ are obtained.
This is one single iteration and it will be iterated multiple times with the nonlinear optimization algorithms in SIMSOPT, e.g. the default trust region reflective algorithm.
After the final plasma equilibrium is obtained, FAMUS is used to optimize the PM layouts.

\section{Numerical results}
\label{section:Numerical}
With the new quasi-single-stage optimization method, we apply it to the MUSE stellarator \cite{Qian2022Simpler}.
MUSE has a major radius of $R$ = 0.3 m and an average toroidal field $B_T$ = 0.15 T.
MUSE was optimized using a typical two-stage optimization approach.
In stage 1, the equilibrium of MUSE was optimized to be quasi-axisymmetric using STELLOPT \cite{stellopt}.
The effective magnetic ripple $\epsilon^{3/2}_{eff}$ \cite{Nemov1999Evaluation} was two orders lower than existing experiments. 
In stage 2, the PM layout was optimized using FAMUS to produce the target equilibrium together with 16 TF coils. 
The magnets were aligned with a toroidal grid and 14 layers of standardized rectangular magnets were placed on a circular cross-sectional torus.
The magnetization orientations were restricted to be normal to the toroidal surface, either straight in or straight out.
FAMUS optimized the distribution of magnets.
Together with some discrete refinements, 12674 magnets per half-period were used (12.7\% of the total available volume).
The difficulties of two-stage optimization also arise in MUSE.
The flux surface produced by the PMs and TF coils did not exactly match the target equilibrium.
A sizable island appeared outside the inner surface of the vacuum vessel.

In our work, we are going to improve the quasi-symmetry, avoid unmatching between the free-boundary equilibrium and the original target, and reduce the number of PMs.
In the quasi-single-stage optimization approach, we started from the MUSE plasma boundary.
The weight $\omega_i$ for each objective function $f_i$ was chosen to satisfy $\omega_if_i=1$.
We set the maximum iterations to 1000 and the output equilibrium did not match all the criteria.
We restarted the optimization with a new initial guess, which is the output of the previous optimization, and adjusted the weights.
The restart was repeated four times before the new PM stellarator design was obtained.
The entire optimization process took about 12 hours in wall-clock time with 16 CPUs and 16 threads per CPU.
We call the new PM equilibrium \mbox{MUSE++}, while the two `+' in `\mbox{MUSE++}' represent the improvement in both the quasi-symmetry and the equilibrium's PMs arrangement.  

\subsection{Main parameters of MUSE++}
\begin{figure}[htbp]
    \centering
    \includegraphics[scale=0.6]{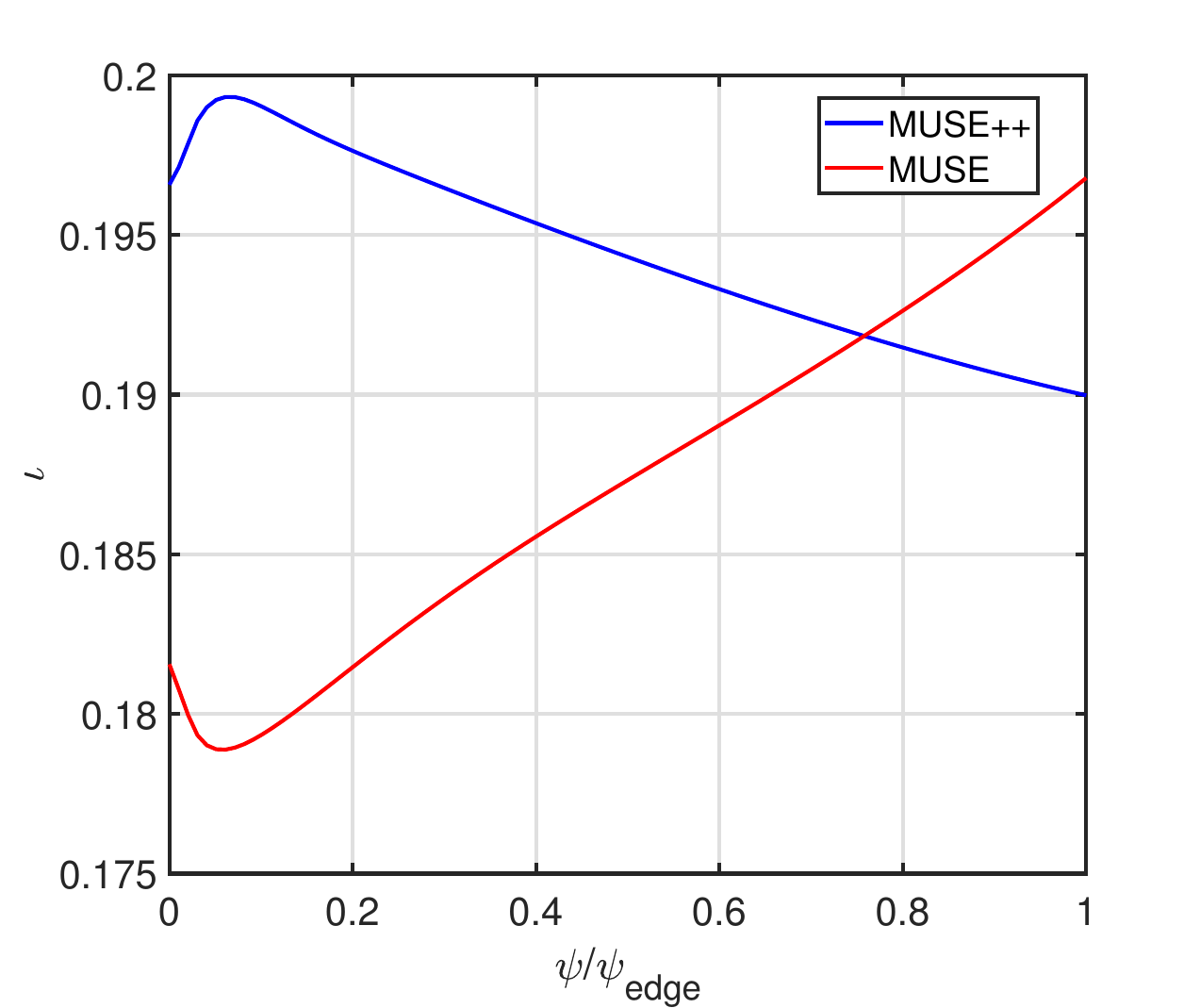}
    \caption{Profiles of rotational transform $\iota$ versus normalized flux label $\psi/\psi_{edge}$.}
    \label{fig:iota}
\end{figure}

\begin{figure}[htbp]
    \centering
    \includegraphics[scale=0.6]{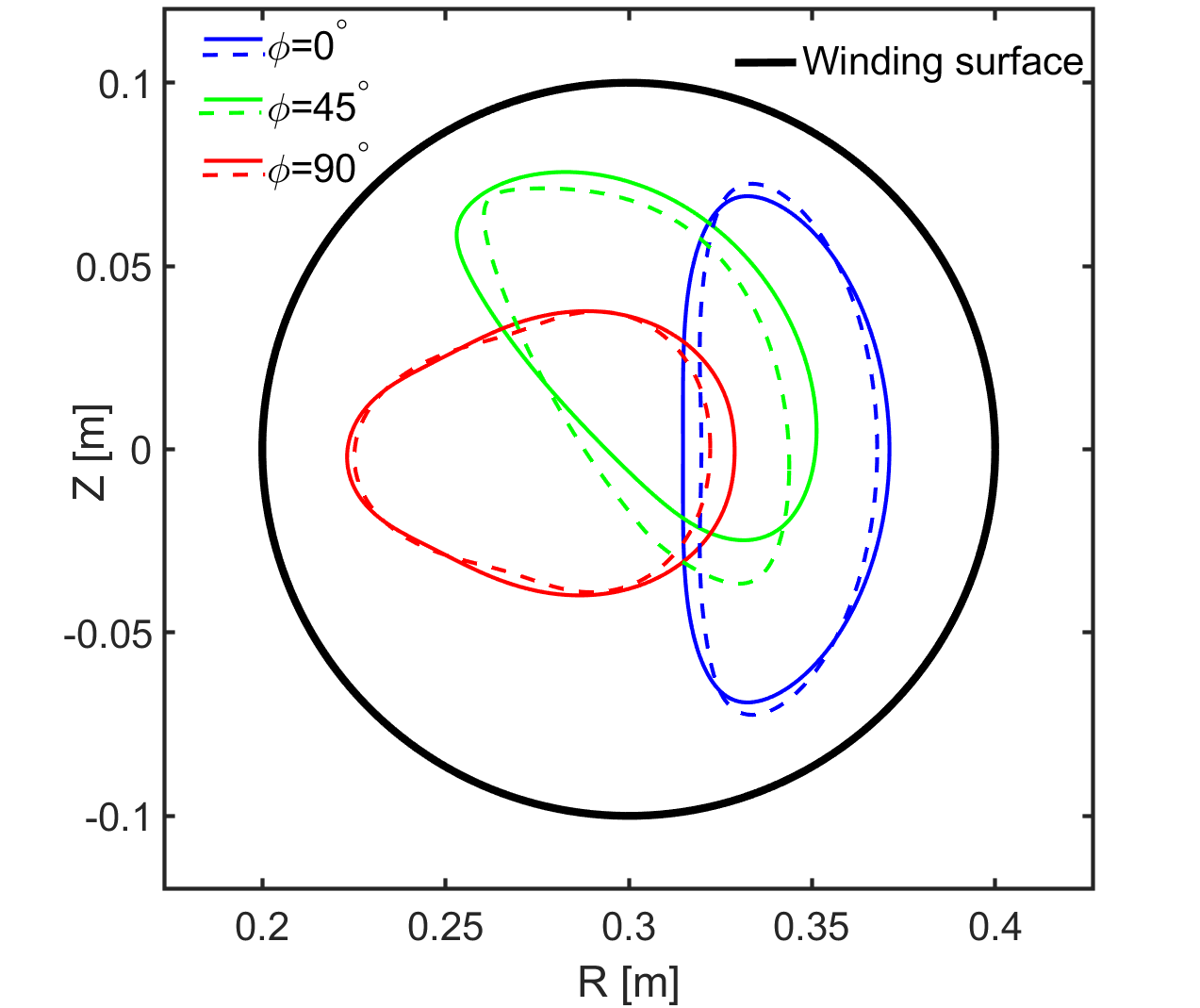}
    \caption{Cross-sections of the last-closed-flux surfaces from VMEC equilibria. Dashed lines for MUSE and solid lines for \mbox{MUSE++}. Three toroidal angles, $\phi=0^\circ$, $45^\circ$, $90^\circ$, are plotted. The winding surface used in REGCOIL is also included.}
    \label{fig:flux_cut}
\end{figure}

Table~\ref{tabel_1} lists the basic equilibrium parameters of MUSE and \mbox{MUSE++}, respectively. It shows that the average rotational transform $\Bar{\iota}$ of MUSE is increased by no more than $6\%$ and the rotational transform profiles are shown in figure~\ref{fig:iota}. The rotational transform iota profiles are not exactly the same, since we are only targeting the average rotational transform, while MUSE was designed to have a small amount of reversed shear. The volume of the boundary surface increases $10\%$ and, meanwhile, the aspect ratio decreases $4\%$ with increasing radius. 
Figure~\ref{fig:flux_cut} shows the cross-sections of the flux surfaces for MUSE and \mbox{MUSE++}. 
The flux surfaces of \mbox{MUSE++} are similar to those of MUSE, since we used MUSE as the initial guess and fixed the major radius and constrained the plasma volume.

\begin{table}
\caption{Basic equilibrium parameters of MUSE and \mbox{MUSE++}.}
\label{tabel_1}
\footnotesize\rm
\begin{tabular*}{\textwidth}{@{}l*{15}{@{\extracolsep{0pt plus12pt}}l}}
\br
Equilibrium & MUSE& MUSE++ \\
\mr
\verb"Number of toroidal field periodicity"/$n_{fp}$&2 &2\\
\verb"Aspect ratio"/$Ap$&7.25 &6.94\\
\verb"Major radius"/$R_0$ ($m$)&0.308 &0.308 \\
\verb"Minor radius"/$a$ ($m$)&0.042 &0.044\\
\verb"Volume"/$V$ ($m^3$) &0.011 &0.012\\
\verb"Average rotational transform"/$\Bar{\iota}$&0.190 &0.195\\
\br
\end{tabular*} 
\end{table}

\subsection{Improvement of quasi-symmetry }
\begin{figure}[htbp]
    \centering
    \includegraphics[width=17cm]{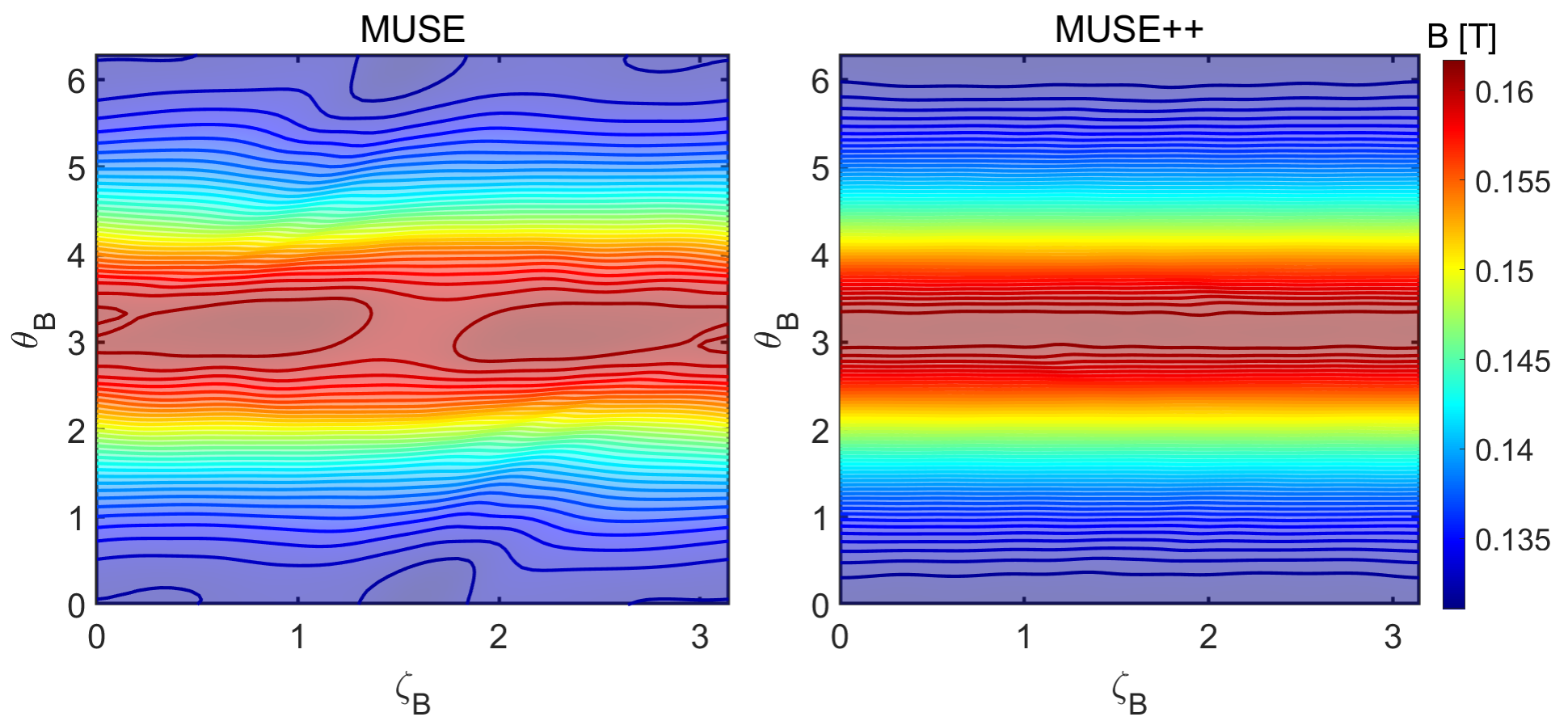}    
\caption{Contours of B on the last-closed-flux surfaces in boozer coordinate. }
\label{fig:B_boozer}
\end{figure}

\begin{figure}[htbp]
 \centering
    \subfloat[$||b_{mn}||_2$]{  
    \includegraphics[width=8.5cm]{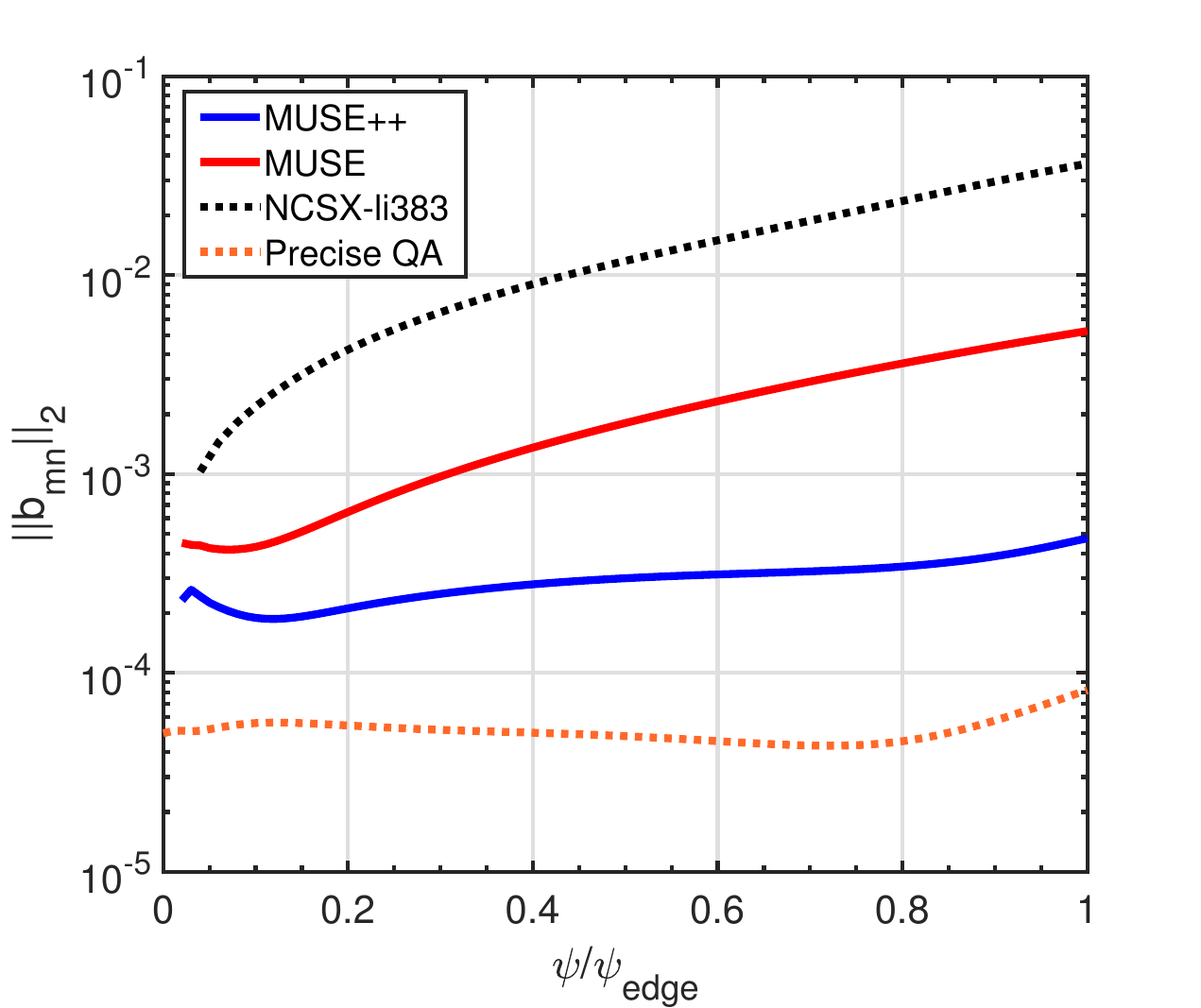}    
    \label{fig:bmn2}
    }
    \subfloat[$\epsilon_{eff}^{3/2}$]{
      \includegraphics[width=8.5cm]{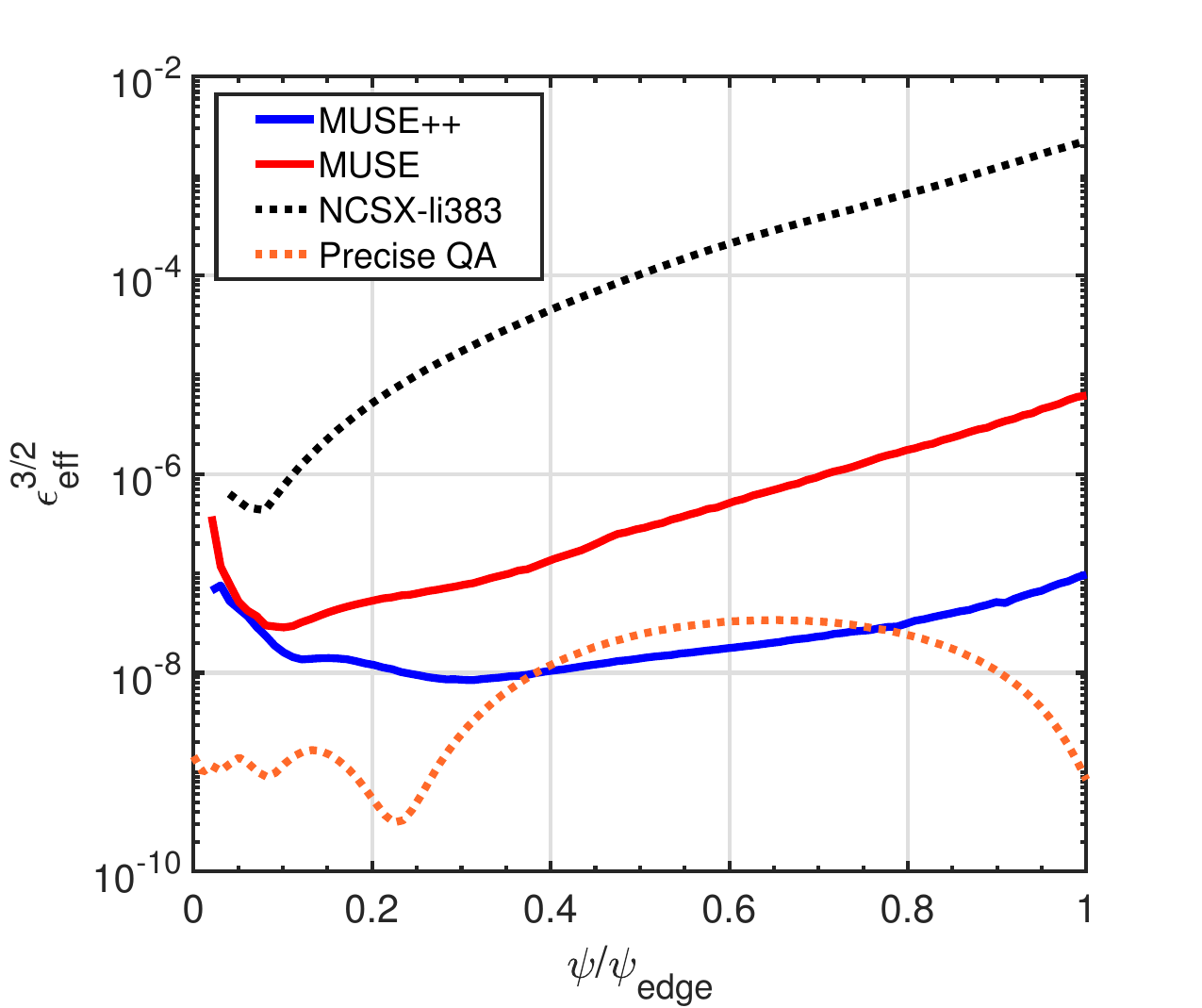}
      \label{fig:epsilon}
    }
\caption{Profiles of the effective helical ripple $\epsilon_{eff}^{3/2}$ and symmetric breaking modes $||b_{mn}||_2$. The NCSX equilibrium is LI383 and the precise QA configuration is from Ref. \cite{Landreman2022PRL} (QA without magnetic well). All the equilibria are from fixed-boundary VMEC runs.}
\label{fig:neoclassical}
\end{figure}

The contours of B on the boundary surfaces are shown in figure~\ref{fig:B_boozer}. The B contours of MUSE deviate from straight lines and have isolated local minima or maxima.  Compared to MUSE, the contours of \mbox{MUSE++} are closer to straight lines without any isolated local minima or maxima. 
The deviation from quasi-axisymmetry can be measured quantitatively by all symmetry-breaking modes as
\begin{equation}
||b_{mn}||_2= \frac{\sqrt{\sum_{m,n\neq0}B_{m,n}^2}}{B_{0,0}} \ ,
\end{equation}
where $B_{m,n}$ is the Fourier modes in Boozer coordinates. $m$ and $n$ are poloidal and toroidal mode numbers. 
In figure~\ref{fig:neoclassical}, we plot the profiles of $||b_{mn}||_2$ and the effective ripple $\epsilon_{eff}^{3/2}$ calculated using the NEO code \cite{Nemov1999Evaluation}.
As a reference, we also include the LI383 fixed-boundary target equilibrium for NCSX.
The $||b_{mn}||_2$ of \mbox{MUSE++} is one order lower than MUSE and two orders lower than NCSX at the edge.
The $\epsilon_{eff}^{3/2}$, which is proportional to the $1/\nu$ neoclassical transport in stellarators, of \mbox{MUSE++} is two orders lower than MUSE and more than four orders lower than NCSX at the edge.
The quasi-symmetry and effective ripple of \mbox{MUSE++} are close to the so-called ``precise quasi-axisymetry'' \cite{Landreman2022PRL}.

\subsection{Improvement of the normal field error and maximum current potential}
\begin{figure}[htbp]
 \centering
    \subfloat[$\chi_B^2$ versus $\chi^2_K$]{  
    \includegraphics[width=8.5cm]{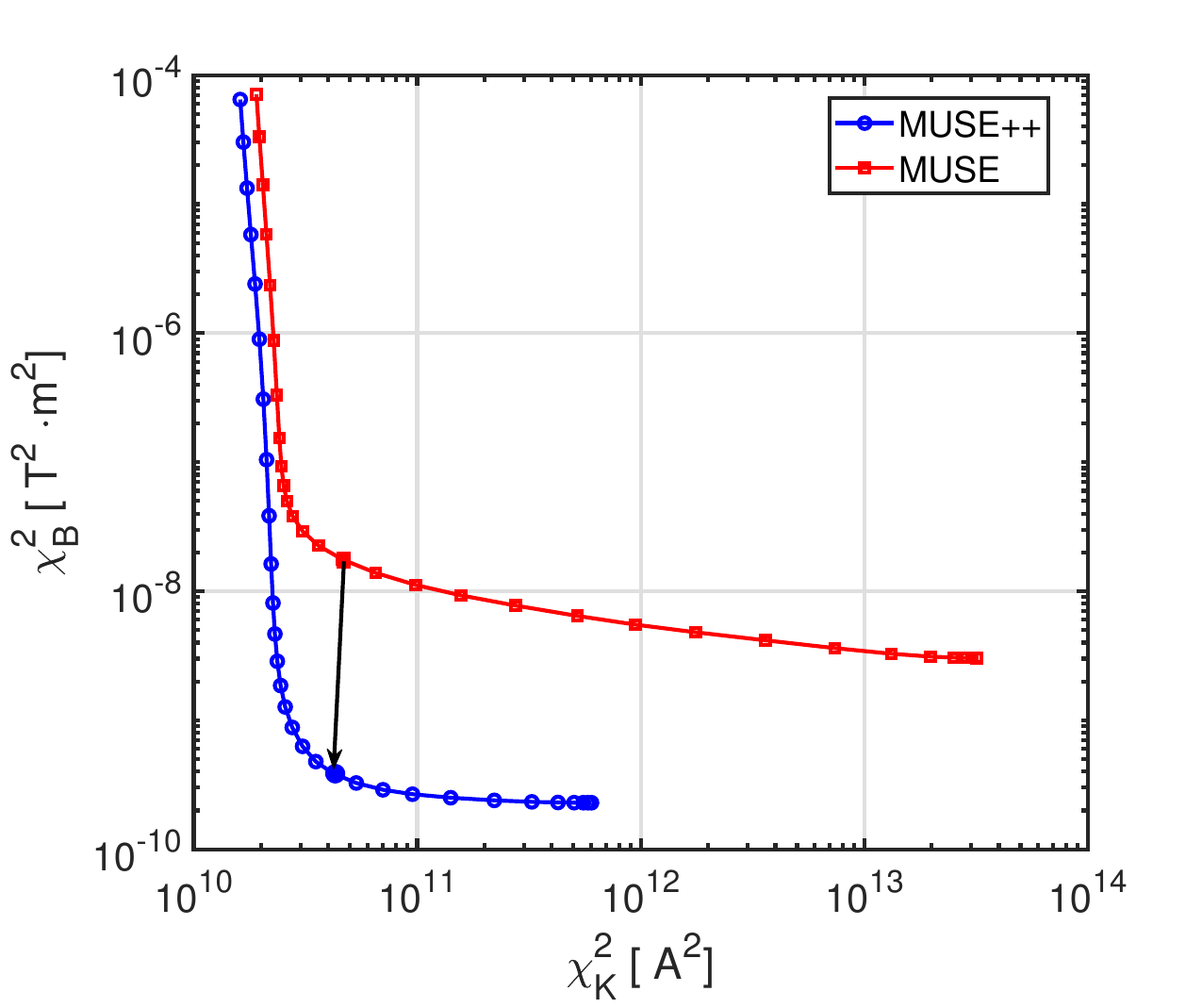}    
    \label{fig:chi2B_chi2K}
    }
    \subfloat[$\chi_B^2$ versus $\Phi_{max}^*$]{
      \includegraphics[width=8.5cm]{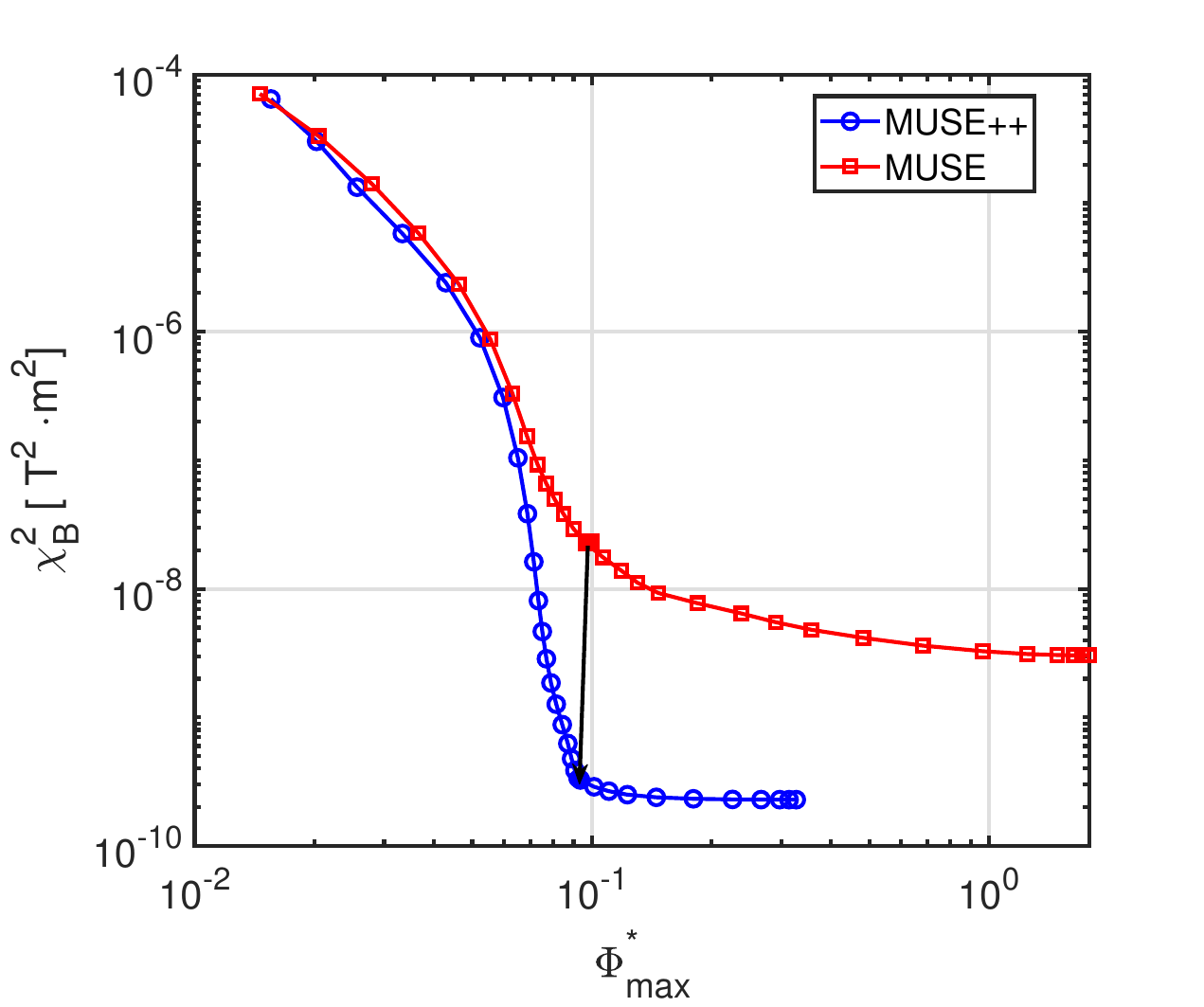}
      \label{fig:chi2B_phimax}
    }
\caption{ Profiles of $\chi_B^2$ versus $\chi^2_K$ (a) and $\Phi_{max}^*$ (b) for the MUSE and \mbox{MUSE++} equilibria. Solid points mark the turning points that minimize $\chi^2_B \cdot \chi^2_K$. The black arrow shows the improvement in turning point from MUSE to \mbox{MUSE++}.}
\label{fig:chi2B_chi2K_phimax}
\end{figure}

In REGCOIL, the curves of the normal field error $\chi_B^2$ and the squared current density $\chi^2_K$ for different regularization parameters $\lambda$ are usually in `L' shapes and yield the so-called `Pareto frontier’ \cite{Landreman2017An}.
The turning points of these `L' curves are considered to be optimal solutions, which correspond to a good balance in both $\chi_B^2$ and $\chi^2_K$. During each optimization iteration, the turning point was found and used to calculate the objective functions of $\chi_B^2$ and $\phi_{max}$. In this work, we select the point with minimum  $\chi_B^2\cdot\chi^2_K$ as the turning point. Figure~\ref{fig:chi2B_chi2K} shows the curves of $\chi_B^2$ versus $\chi^2_K$ for MUSE and \mbox{MUSE++}. The turning points with minimum $\chi_B^2\cdot\chi^2_K$ are marked on the two curves (solid points). We can see that the normal field error of \mbox{MUSE++} is more than one order lower than that of MUSE at the turning point. This improvement will reduce the mismatch between the free-boundary equilibrium and the target. Another optimization target, $\Phi_{max}^*$, which is the maximum current potential normalized to the total poloidally-linked current (currents in TF coils), is also decreased by $15\%$, as shown in Figure~\ref{fig:chi2B_phimax}. The metric reduces the number of permanent magnets required.

\section{Improved PM design using FAMUS}
\label{section:Improved}
In this section, we use FAMUS \cite{Zhu2020Topology} to design practical layouts for permanent magnets.
MUSE PM design was performed with FAMUS together with discrete refinements. 
The MUSE team has made a tremendous effort to improve the final design with many refinement techniques. Details are documented in Ref. \cite{Qian2023Design}.
Here, we are not going to adopt the same procedure, as it is time-consuming and requires a lot of manual refinements.
The problem of PM optimization is ill-posed and there exist numerous solutions. 
Our main goal is to validate that we can obtain improved PM designs under the same conditions when
the plasma equilibrium changes from MUSE to \mbox{MUSE++}. 

\begin{figure}[htbp]
    \centering
    \includegraphics[scale=0.6]{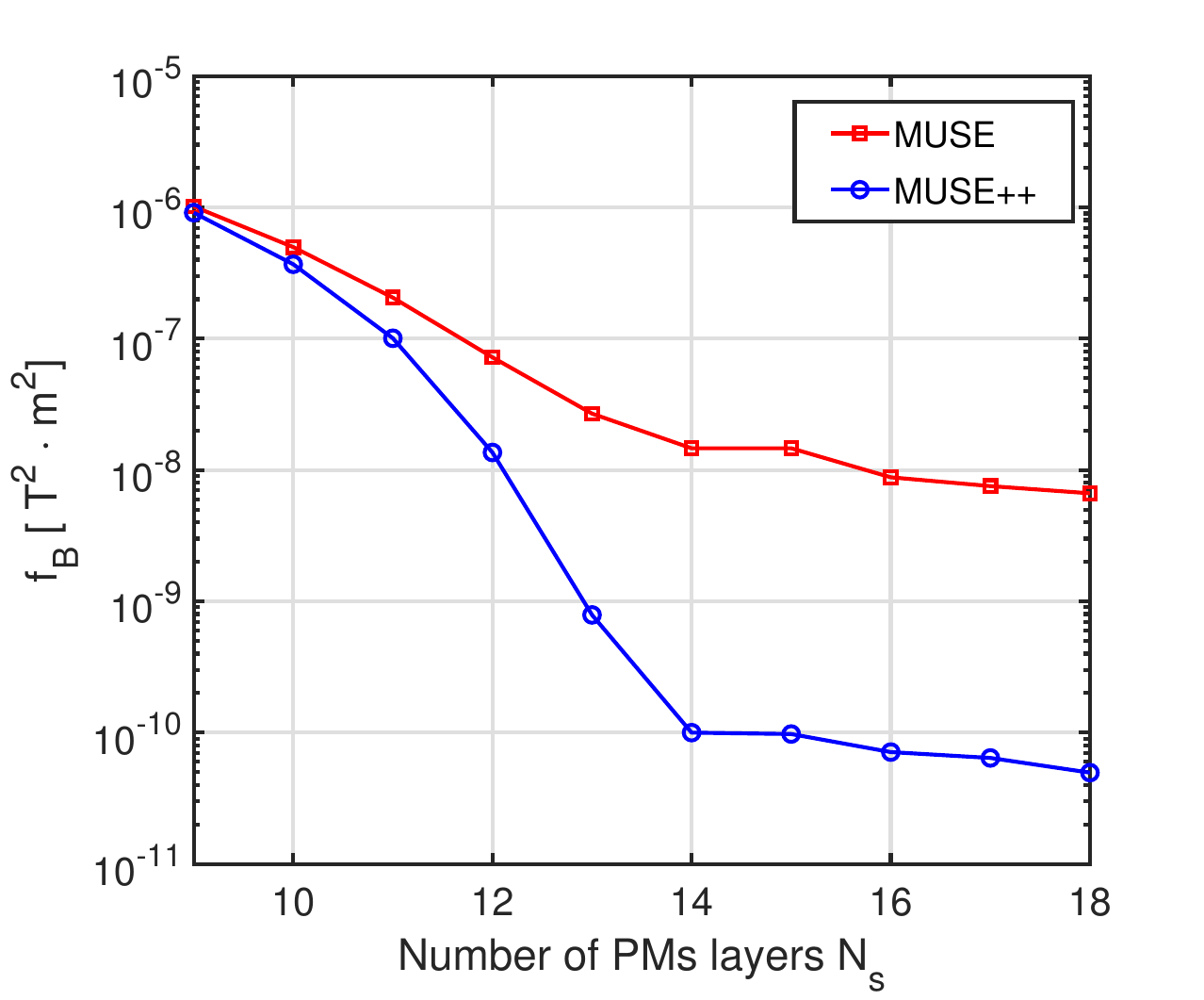}
\caption{The relationship between the achieved normal field error in FAMUS and the number of PM layers for both \mbox{MUSE} and \mbox{MUSE++} equilibria.}
\label{fig:chi2b_layers}
\end{figure}

We define a simple procedure to optimize PMs. The geometry of permanent magnets and TF coils are the same as MUSE. All the dipoles are placed on a circular torus and aligned in a toroidal grid.
The size of each magnet is $0.25 \times 0.25 \times 0.0625 \  \mathrm{inch}$ and is represented as an ideal dipole. 
There are in total 14 layers and the innermost layer is slightly outside the winding surface that is used in the one-stage optimization.
The number of PM layers is related to the achieved normal field error.
As shown in figure \ref{fig:chi2b_layers}, for both \mbox{MUSE} and \mbox{MUSE++}, 14 layers are sufficient to produce the required magnetic field.
Note that \mbox{MUSE++} generally has a lower normal field error when using the same number of layers.
In other words, \mbox{MUSE++} needs fewer layers when pursuing the same level of normal field errors.
This reduction comes from the PM metric we used in the quasi-single-stage optimization.
The total number of PMs is 77196 per half period, and all the magnets are restricted to be in the normal direction, either straight in or straight out. First, we make all dipoles have the same magnetization by setting the normalized density parameter $p=1$ with $q=3$ in equation (\ref{eq:f_D}). Then we run FAMUS to optimize $p$ to minimize the normal field error $f_B$ and the binary penalization $f_D$. After FAMUS optimization, all the dipoles with $|p|<0.1$ are removed from the dipole list. The normalized density parameter $p$ of the remaining dipoles will be rounded to $\pm1$. The new solution will be the initial guess for another FAMUS run. The above step will be repeated several times until it converges.

\begin{figure}[htbp] 
 \centering 
    \subfloat[Inboard view of MUSE-0 ]{  
    \includegraphics[width=8cm]{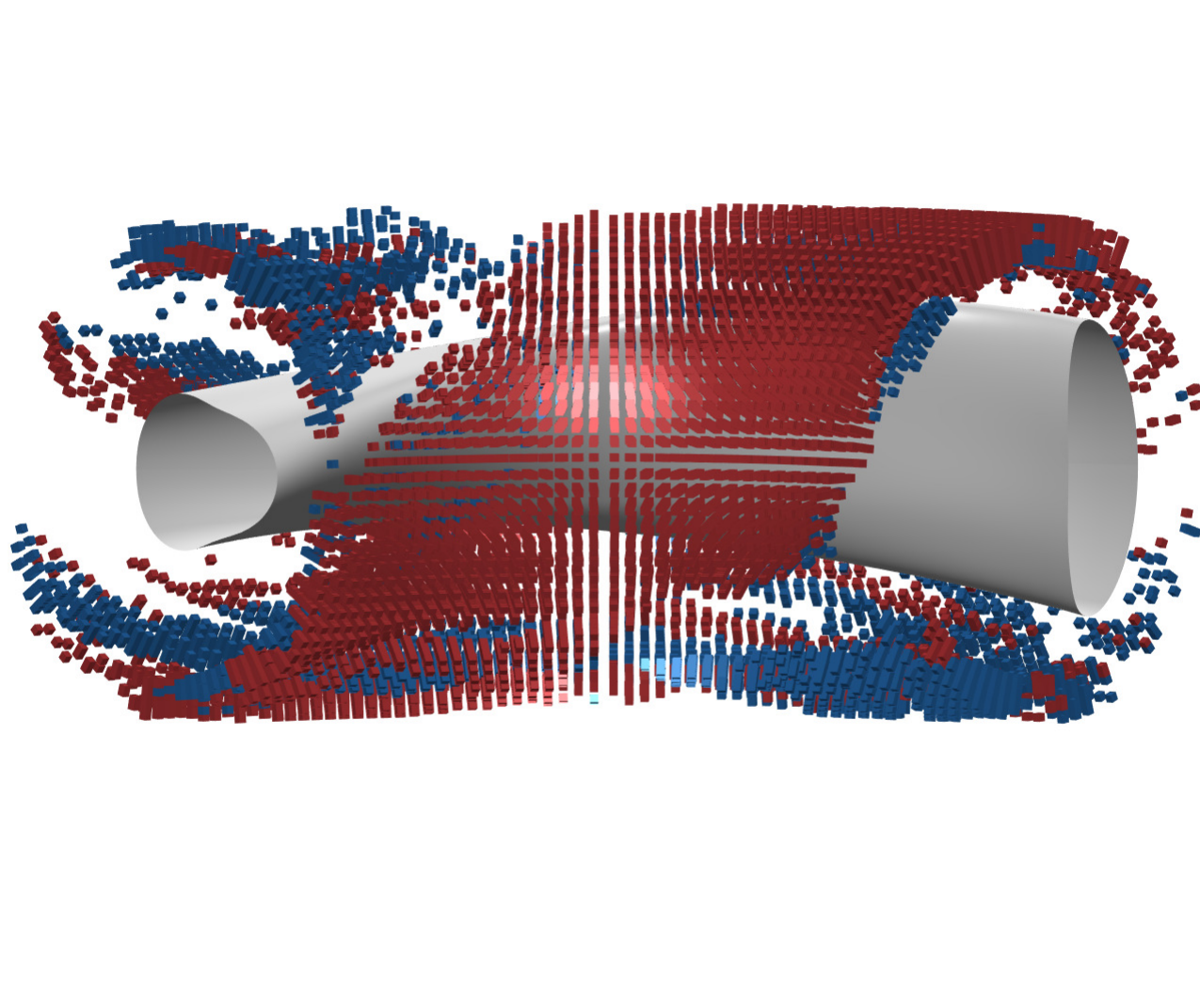}    
    \label{fig:dipoles_muse_inboard}
    }
    \subfloat[Inboard view of MUSE++ ]{  
    \includegraphics[width=8cm]{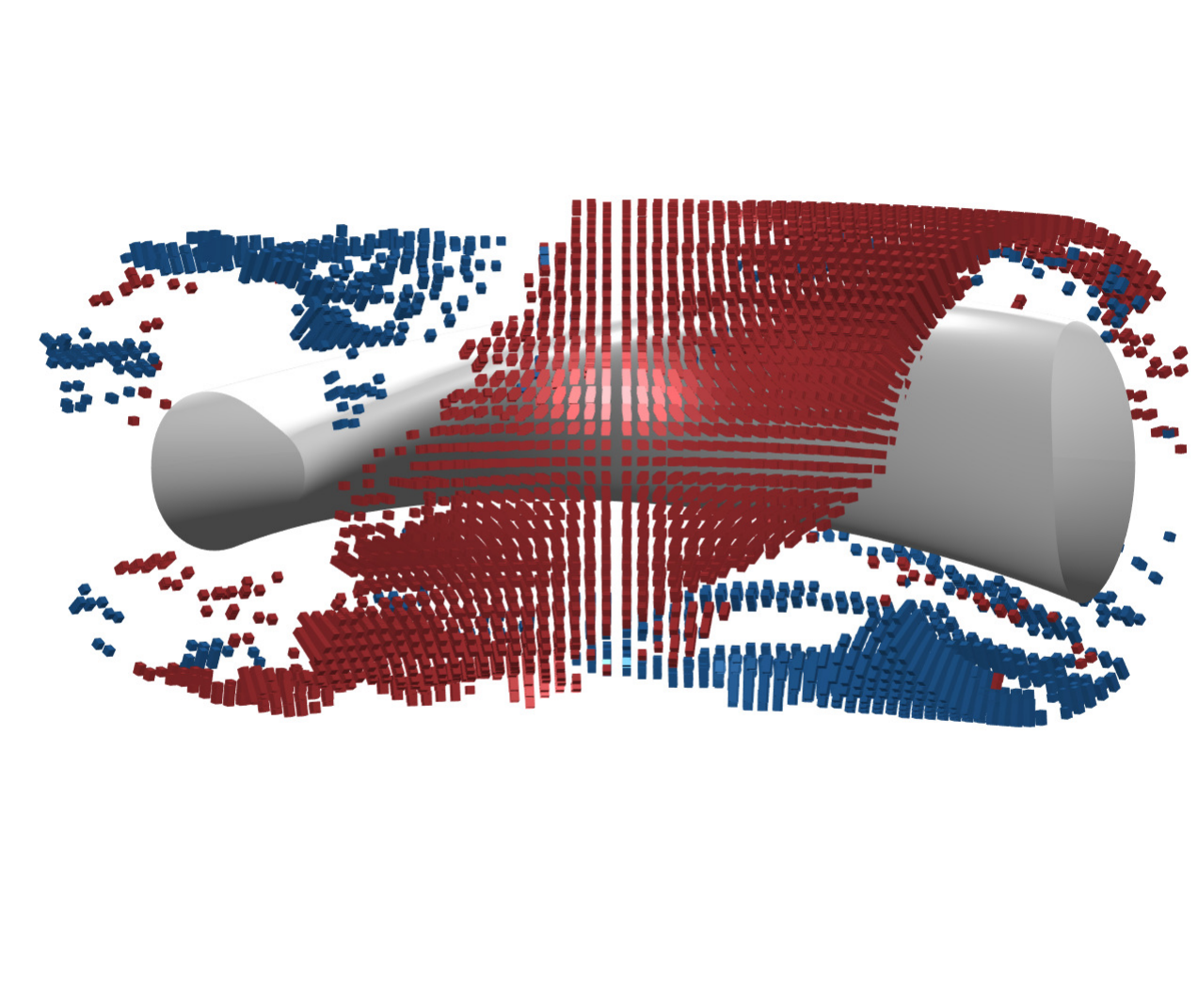}    
    \label{fig:dipoles_muse++_inboard}
    }  

    \subfloat[Outboard view of MUSE-0 ]{  
        \includegraphics[width=8cm]{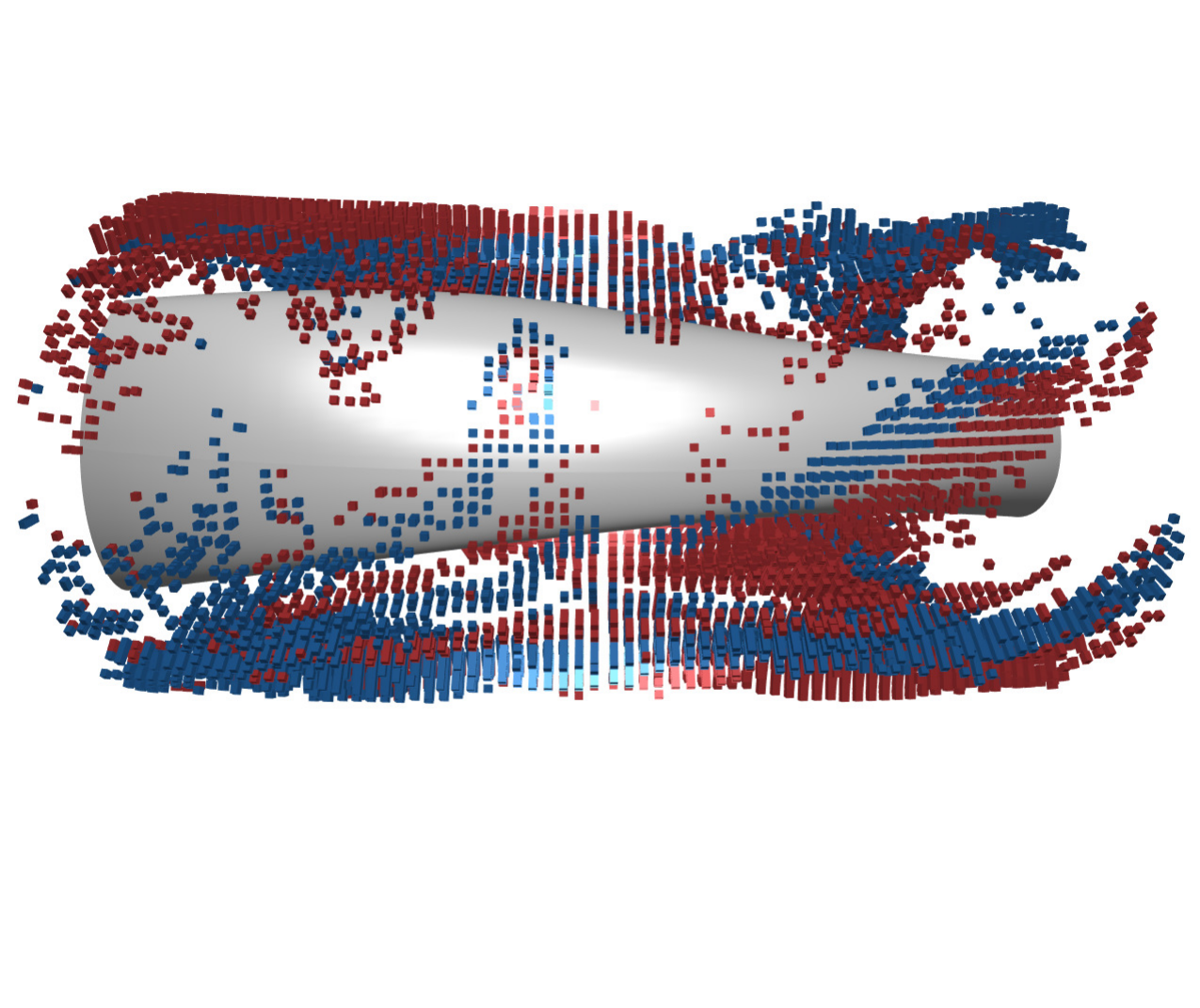}    
        \label{fig:dipoles_muse_outboard}
        }
    \subfloat[Outboard view of MUSE++ ]{  
    \includegraphics[width=8cm]{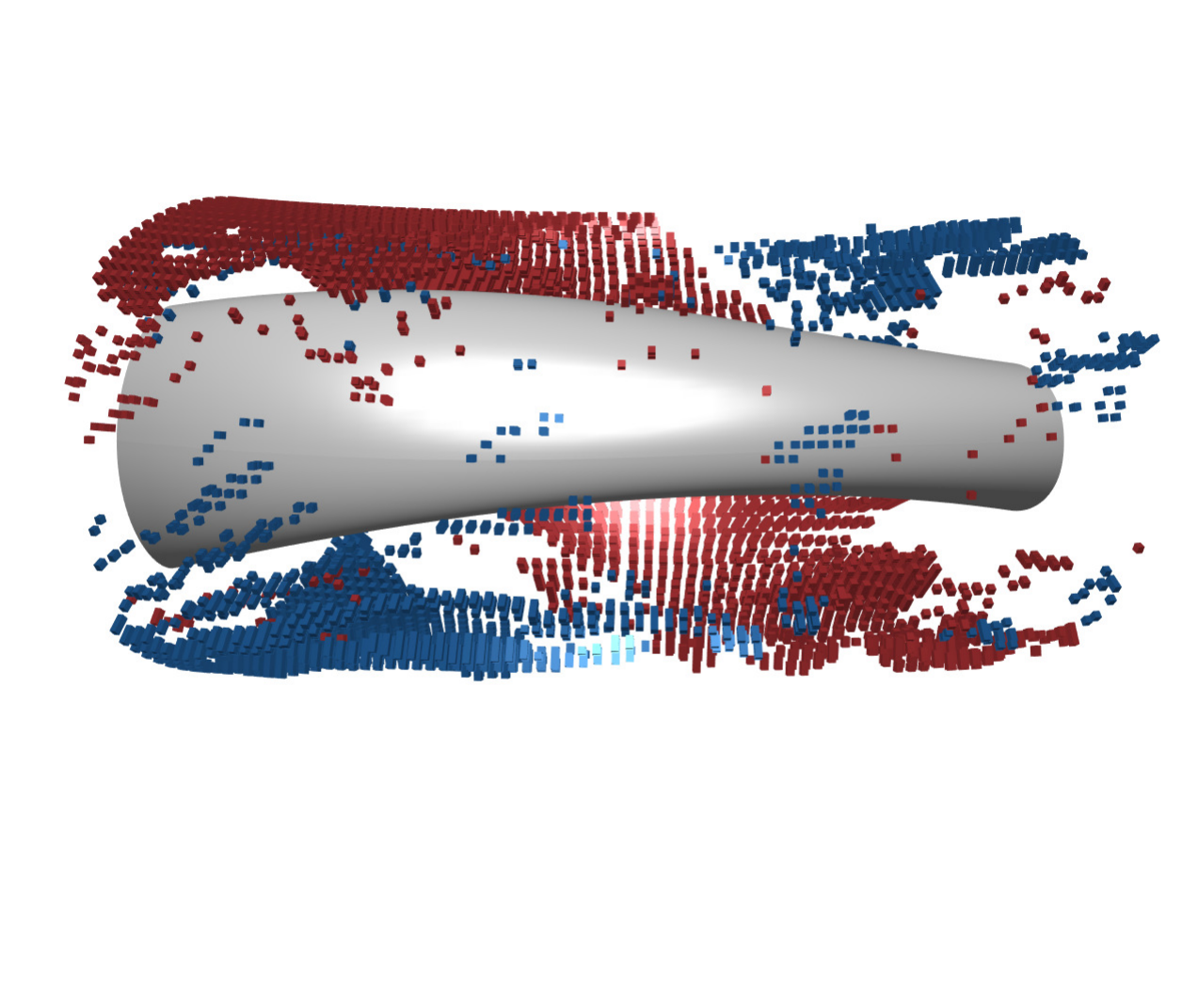}    
    \label{fig:dipoles_muse++_outboard}
        }  
\caption{Inboard and outboard views of the permanent magnets for \mbox{MUSE-0} and \mbox{MUSE++} in the half-period. 
The two different colors of PM refer to the inward and outward orientation of magnetization.}
\label{fig:dipoles_inboard_outboard_view}
\end{figure}

\begin{figure}[htbp]
    \centering
    \includegraphics[width=17cm]{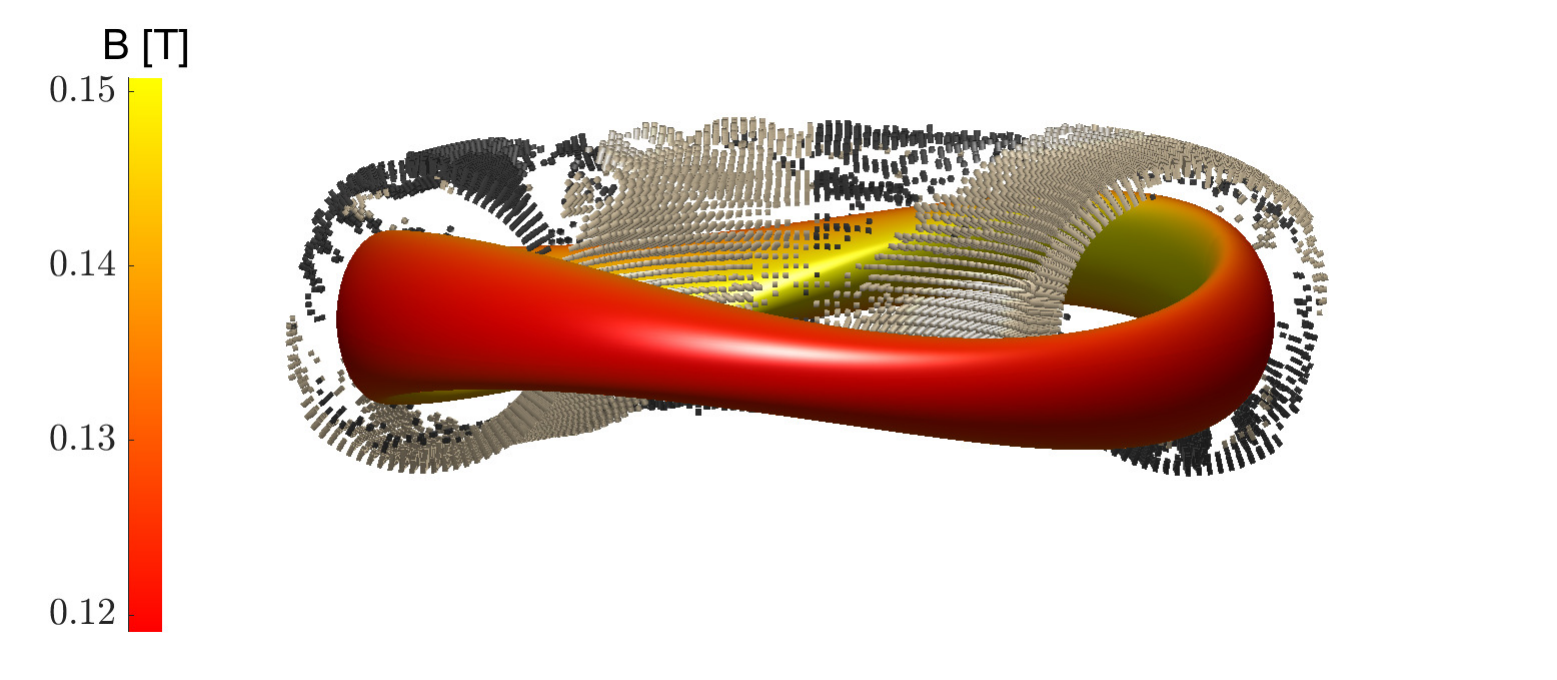}
\caption{3D view of PMs and LCFS for \mbox{MUSE++}.The two different colors of PM refer to the inward and outward orientations. The color on the flux surface indicates the strength of the magnetic field.}
\label{fig:dipoles_muse++}
\end{figure}

We have adopted the above simple procedure to design PMs for MUSE and \mbox{MUSE++}.
The new PM design for MUSE differs from the original MUSE PM design \cite{Qian2023Design} because we are not 
following the same procedure. In addition, the MUSE magnets were designed to avoid specific regions
to allow diagnostic access to the plasma, including vertical views.
To distinguish from the original MUSE PM design, we shall call the new PM design ``\mbox{MUSE-0}'' 
since it is similar to one of the \mbox{generation-0} solutions in figure 7 of Ref. \cite{Qian2023Design}. 
The comparisons of PM among MUSE++, MUSE-0, and MUSE are listed in table \ref{tabel_2}.
Note that MUSE++, MUSE-0, and MUSE denote a specific PM design in this section, while MUSE and MUSE++ denote the plasma equilibrium in section \ref{section:Numerical}.
MUSE refers to the actual PM design for the MUSE stellarator that was carried out with many manual refinements \cite{Qian2023Design}.
MUSE-0 refers to the PM design for the MUSE equilibrium, and it comes from a standard, simplified procedure introduced above.
MUSE++ is the PM design for the MUSE++ equilibrium and it follows the same optimization procedure as MUSE-0.
The \mbox{MUSE-0} design has 21082 magnets per half period (5.40 liters of total magnetized volume).
Following the same procedure, \mbox{MUSE++} only needs 15062 magnets per half period (3.86 liters of 
total magnetized volume), which is about a 28.6\% reduction.
In figure \ref{fig:dipoles_inboard_outboard_view}, we plot the magnets for \mbox{MUSE-0} and \mbox{MUSE++}.
The outboard side of \mbox{MUSE++} is much more empty, which is favorable to provide more access
for heating and diagnosis. 
Figure~\ref{fig:dipoles_muse++} shows the 3D view of the plasma and permanent magnets for \mbox{MUSE++}.
For clarity, TF coils are not shown and PM is only plotted in one period.

\begin{table}
\caption{PM comparisons among MUSE++, MUSE-0 and MUSE.}
\label{tabel_2}
\footnotesize\rm
\begin{tabular*}{\textwidth}{@{}l*{15}{@{\extracolsep{0pt plus12pt}}l}}
\br
PM design    & MUSE++& MUSE-0& MUSE\\
\mr
Number of PM (per half period) &15062 &21082 &12674\\
Magnetized volume (full torus) ($L$) &3.86 &5.40 &3.24\\
Fraction of MUSE++ &$100.0\%$ &$140.0\%$ &$84.1\%$ \\
$f_B$ ($T^2 \cdot m^2$) & $1.52\times10^{-9}$ & $1.07\times10^{-8}$ &$1.29\times10^{-8}$\\

\br
\end{tabular*} 
\end{table}

\begin{figure}[htbp]
    \centering
    \includegraphics[scale=0.6]{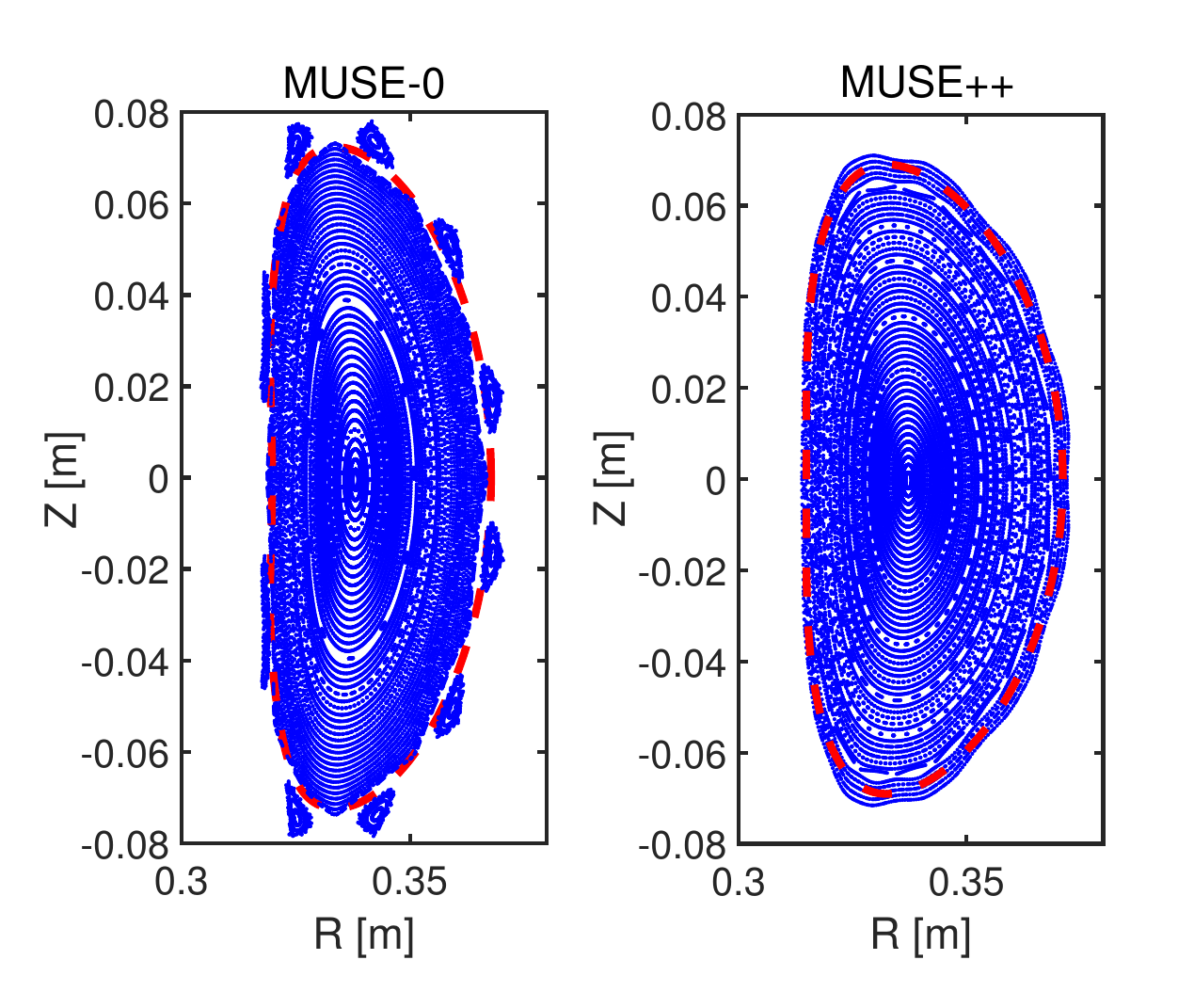}
\caption{Poincar\'{e} plots from TF coils and permanent magnets at $\phi=0^\circ$ plane. Red points are the target boundaries and blue points are field lines from TF coils and permanent magnets.}
\label{fig:Poincare_plot}
\end{figure}

\begin{figure}[htbp]
    \centering
       \subfloat[$||b_{mn}||_2$]{  
       \includegraphics[width=8.5cm]{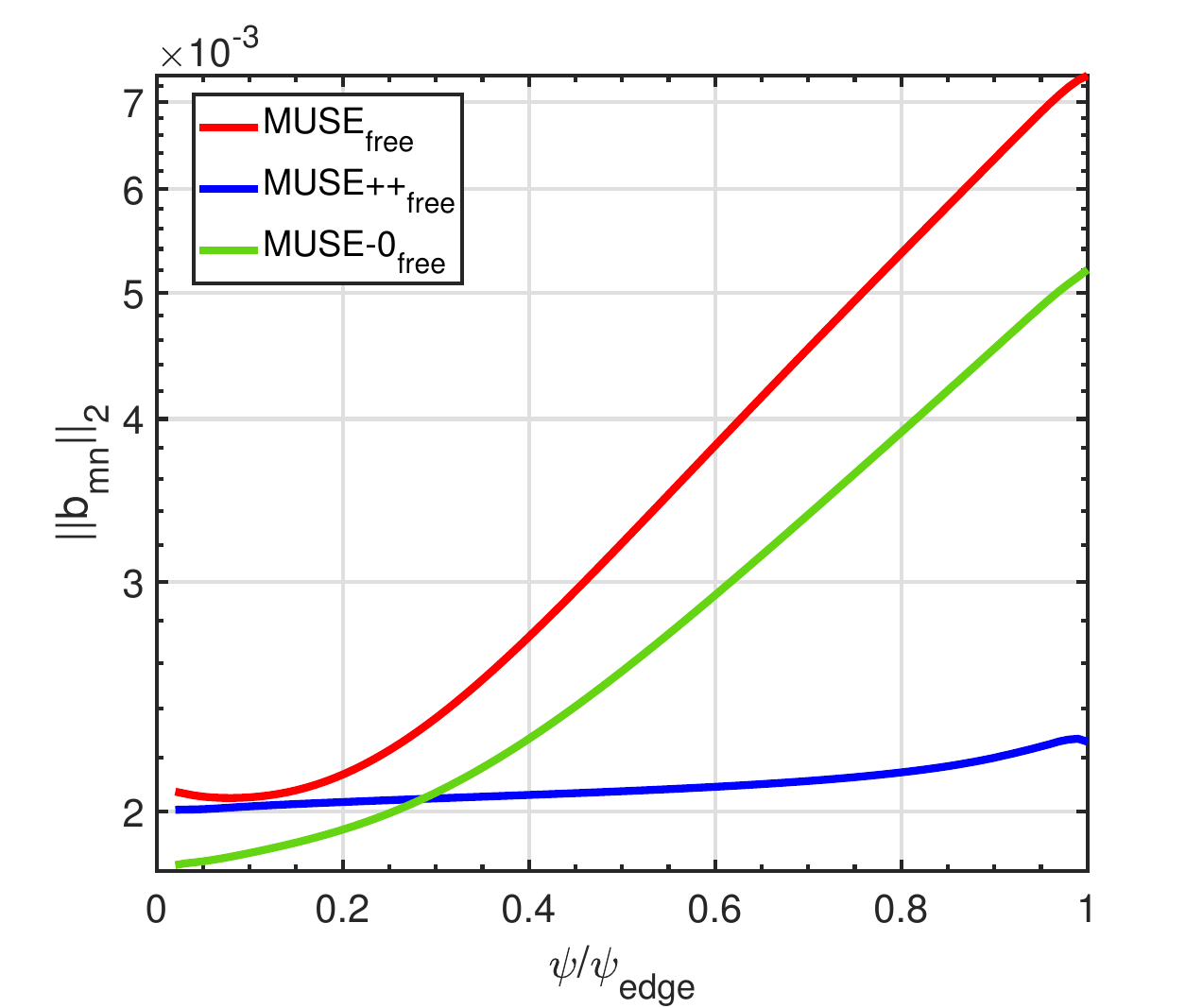}    
       \label{fig:bmn2_free}
       }
       \subfloat[$\epsilon_{eff}^{3/2}$]{
         \includegraphics[width=8.5cm]{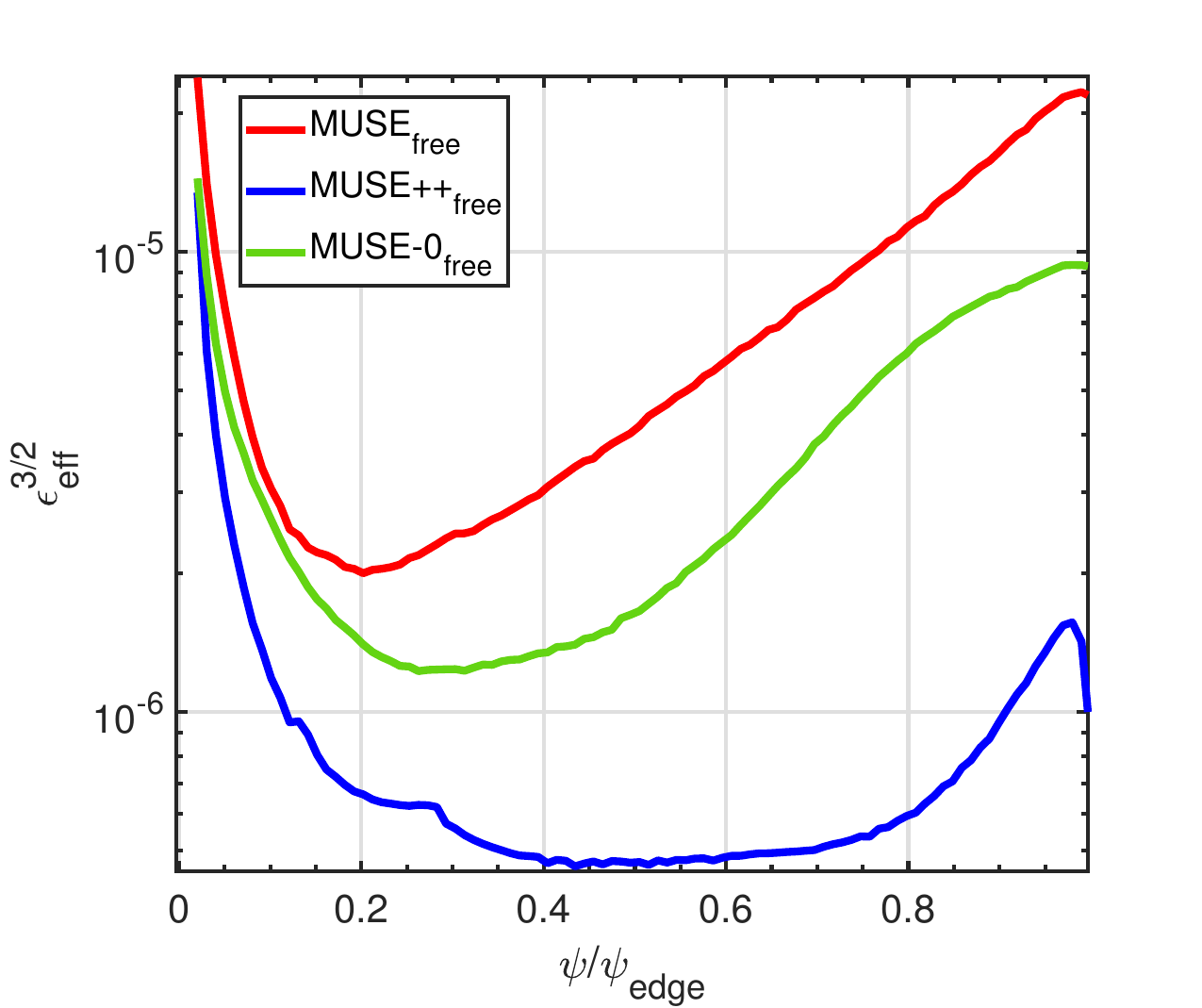}
         \label{fig:epsilon_free}
       } 
   \caption{Profiles of the effective helical ripple $\epsilon_{eff}^{3/2}$ and symmetric breaking modes $||b_{mn}||_2$. 
   All the equilibria are from free-boundary VMEC runs. 
   The vacuum magnetic fields are from the discrete PM layouts in Table~\ref{tabel_2} 
   together with 16 TF coils.  }
   \label{fig:neoclassical_free}
   \end{figure}

In addition to the reduction of permanent magnets, the normal field error $f_B$ of \mbox{MUSE++} is much smaller than that of \mbox{MUSE-0}.
$f_B$ is $1.07\times 10^{-8}$ $\mathrm{T^2 \cdot m^2}$ for \mbox{MUSE-0} and $1.52\times 10^{-9}$ $\mathrm{T^2 \cdot m^2}$ for \mbox{MUSE++}. 
The improvement in the normal field error brings better matches in the free-boundary equilibrium.
Figure~\ref{fig:Poincare_plot} shows the Poincar\'{e} plots from the TF coils and permanent magnets.
There exists an island chain at the edge for \mbox{MUSE-0}, which is similar to the original MUSE design.
While in \mbox{MUSE++}, the Poincar\'{e} plots better match the target flux surface, and the islands disappear.
The results indicate that the improvement in the normal field error works. Moreover, the free boundary equilibrium 
of MUSE++ also reveals better quasi-symmetric than MUSE and MUSE-0 which is shown in Figure~\ref{fig:neoclassical_free}.

\section{Discussion and conclusions}
\label{section:Discussion}
The paper introduces a quasi-single-stage optimization method for designing permanent magnet stellarators. 
This new approach employs straightforward metrics to penalize permanent magnet (PM) layouts and integrates
them with conventional fixed-boundary optimization targets. 
The method is implemented within the SIMSOPT framework, incorporating REGCOIL.

We applied this new quasi-single-stage optimization method to enhance the MUSE design, resulting in the
discovery of a new quasi-axisymmetric equilibrium. 
The new configuration, referred to as \mbox{MUSE++}, exhibits an order-of-magnitude improvement in
quasi-axisymmetry and a one-order reduction in normal field error compared to MUSE.
We used FAMUS to design realistic PM layouts for both the \mbox{MUSE++} and MUSE equilibria. 
MUSE++ is both a new equilibrium and one of its magnet solutions. 
MUSE-0 is a PM solution to the original MUSE equilibrium using the same design procedure as \mbox{MUSE++}. 
As a PM solution, \mbox{MUSE++} uses 28.6\% fewer magnets than \mbox{MUSE-0}. 
However, both use more magnets than the MUSE PM solution that was constructed. 
This is because the original MUSE design pursued further refinement techniques that were not used in this paper.
We could apply those techniques for MUSE++, but that work is beyond the scope of this paper.
The PM designs considered in this paper also do not exclude ports, while the actual MUSE does.

In future work, we will explore the application of our quasi-single-stage optimization method to other PM 
stellarator designs, such as PM4Stell. Additionally, we aim to incorporate more physical objectives, 
such as MHD stability, turbulence transport, alpha particle confinement, etc. Furthermore, 
our approach can be extended to optimize stellarators with coils.

\section{Acknowledgement}
We sincerely thank the MUSE team for sharing the data and Dr. M. Zarnstorff for fruitful discussions.
We would like to acknowledge the code development teams of SIMSOPT, REGCOIL, VMEC, and FAMUS. 
This work was supported by the National Natural Science Foundation of China and the Anhui Provincial Key Research and Development Project with grant number 2023a05020008.
Numerical calculations were performed at the Hefei Advanced Computing Center.

\newcommand{\newblock}{}
\bibliographystyle{unsrt}
\bibliography{ref}

\end{document}